\documentclass{webofc}

\usepackage[varg]{txfonts}   
\usepackage{hyperref}
\usepackage{url}
\usepackage{graphicx}
\usepackage[figuresright]{rotating}
\usepackage{makecell}
\usepackage{bm,amsmath,amssymb}
\usepackage[mathscr]{eucal}
\usepackage{multirow}
\usepackage{xcolor}
\usepackage{mathrsfs}
\usepackage{mathtools}
\usepackage{slashed}
\usepackage{graphics}
\usepackage{graphicx}
\usepackage{subfigure}
\usepackage{dsfont}
\usepackage{longtable}
\usepackage{bbm} 
\usepackage{float}
\usepackage{tcolorbox}
\usepackage{lipsum}
\usepackage{cancel}
\usepackage{enumerate}
\usepackage{diagbox}
\hypersetup{colorlinks=true,citecolor=blue,urlcolor=blue,linkcolor=blue}
\newcommand{\rmd}{{\rm{d}} }
\begin{document}
\title{Jet Energy-Energy Correlator in Cold QCD Matter}
\author{
        \firstname{Yu} \lastname{Fu}\inst{1}\footnotemark[1] \and
        \firstname{Berndt} \lastname{M\"uller}\inst{1}  \and
        \firstname{Chathuranga} \lastname{Sirimanna}\inst{1}\fnsep\footnotemark[1]
}

\institute{Department of Physics, Duke University, Durham, NC 27708, USA}

\abstract{We present a study of medium-induced modifications to the energy-energy correlator (EEC) for jets in cold nuclear matter. For electron–nucleus collisions, at leading order in the QCD coupling and in the jet–medium interaction, we derive an analytic expression for the EEC modification as a function of the opening angle and show that the modification is strongest at large angles within the jet cone. The dependence on jet energy, the transport properties of cold nuclear matter, and the in-medium path length is made explicit. We further extend the analysis to gluon jets in proton–nucleus collisions and compare with preliminary proton–lead data from the LHC. Possible effects of comovers on the EEC in proton–nucleus collisions are also discussed.

}
\maketitle
\renewcommand{\thefootnote}{\fnsymbol{footnote}}
\footnotetext[1]{speakers}
\renewcommand{\thefootnote}{\arabic{footnote}}
\section{Introduction}
\label{intro}
The energy-energy correlator (EEC) is a well-established observable studying quantum chromodynamics (QCD) in a variety of collider experiments \cite{Moult:2025nhu}. It measures the total energy deposited in two ideal detectors as
a function of the angle between the detectors. Modern high-energy colliders with the exceptional angular resolution of tracking detectors and calorimeters have enabled the precise study of EECs inside jets, which is an infrared and collinear safe jet substructure observable quantifying the detailed structure of energy flow within jets. The relevant studies about EEC open new opportunities to deepen our understanding of QCD phenomena in vacuum and nuclear medium, ranging from imaging the confining transition from quarks and gluons to hadrons \cite{Komiske:2022enw,ALICE:2024dfl}, to probing the scale dependence of the quark-gluon plasma \cite{Andres:2022ovj,Yang:2023dwc}.

Despite extensive investigations of jet EECs in vacuum and hot media, their behavior in the cold nuclear matter(CNM) remains largely unexplored. Electron-nucleus (e+A) collisions at future facilities like the EIC, as well as the current proton-nucleus (p+A) collisions at the LHC, offer unique opportunities to probe jet interactions with CNM and study the transport properties of the medium. In this proceeding, we outline a recent study of jet EECs in CNM \cite{Fu:2024pic}, based on e+A and p+A collisions formulated in the higher-twist (HT) approach.

\section{Formalism}
\label{sec-CNM}
To explore jet EECs theoretically, we examine how the energy flow correlation $\langle\mathcal{E}_{\vec{n}_i} \mathcal{E}_{\vec{n}_j} \rangle$ within the jet depends on the opening angle $\theta$ of directions $\vec{n}_i$ and $\vec{n}_j$. In the small-angle region, the EEC reflects the distribution of uncorrelated hadrons, while in the large-angle region it reflects the structure of partonic splittings in perturbative QCD. These domains are separated by a broad peak governed by confinement effects, whose position scales with the total jet momentum $p_T$ as $p_T\theta_{\rm peak} \approx 2.4$ GeV/c \cite{ALICE:2024dfl}. We focus on the perturbatively calculable large-angle region where the CNM modifies the partonic splitting pattern inside the jet.
To the LO in the QCD coupling, the energy flow in the jet initiated in a $q\to q+g$ splitting is determined by the momentum and angle distribution of the offspring quark and gluon. We therefore compute the energy-weighted differential cross-section as $\frac{\rmd \Sigma}{\rmd \theta} =\int_0^1 \rmd z \frac{\rmd \sigma_{qg} }{\rmd \theta}z(1-z)$,
where $\sigma_{qg}$ is the inclusive cross-section for a $q\to q+g$ splitting and $z$ is the large momentum fraction carried by the offspring quark. Based on this definition, we discuss the calculation of the energy-weighted cross section in e+A and p+A collisions in the rest of this section.

\subsection{Jet EECs in DIS}
For the DIS in e+A collision, the large size of the nucleus provides a cold nuclear environment for the propagating jets. The nuclear effect can be quantified by comparing the jet propagation without a nuclear background, which is achieved in the e+p collision. 
In the Breit frame, the momentum transfer between the electron and the proton or nucleus is $q = [{Q^2}/{2 q^-}, q^-, 0_\perp]$, where $Q^2$ is the photon's virtuality. Within the collinear factorization framework\cite{Collins:1989gx,Wang:2001ifa}, the energy-weighted differential cross section for DIS can then be expressed as follows,
\begin{align}
    \frac{\rmd \Sigma}{\rmd x_B \rmd Q^2 \rmd\theta} = \sum_q \frac{e_q^2 \alpha^2}{2 Q^2 x_B^2 s^2} L_{\mu \nu} 
    H_0^{\mu \nu} (x_B) f(x_B) \mathcal{K}(\theta),
\end{align}
where $x_B = Q^2/2 p^+ q^-$ with $p^+$ the large light-cone momentum of nucleon, $H_0^{\mu \nu}(x_B)$ is the Born term for hard scattering, $f(x_B)$ is the parton distribution function (PDF), and the kernel $\mathcal{K}(\theta)$ for leading twist (LT) and next-to-leading twist (NLT) are given by,
\begin{align}
    \mathcal{K}_{LT}(\theta) &= \frac{\alpha_s C_F}{2 \pi} \int_0^1 \rmd z \frac{(1-z)z}{\theta} P_{qg}(z), \\
    \mathcal{K}_{NLT}(\theta,q^-) &= \frac{\alpha_s C_A}{2\pi }\int_0^1 \rmd z  \frac{1}{\pi } \frac{8P_{qg}(z)}{\theta^3z(1-z)(q^-)^2} 
     \int_0^L \rmd\xi^- \hat{q}(\xi^-) \sin^2 \left( \frac{z(1-z)q^-\theta^2}{4}\xi^- \right).
     \label{eq::xsection_NLT}
\end{align}

In e+p collisions, only $\mathcal{K}_{LT}$ dominates, while the $\mathcal{K}_{NLT}$ terms are power suppressed. In the presence of medium interactions, however, the relevant kernel is the combination $\mathcal{K}_{LT} + \mathcal{K}_{NLT}$, together with the nuclear parton distribution functions (nPDFs). We note that the angular dependence of the NLT contribution to the energy-weighted differential cross section is linear in $\theta$ when the argument of the sine function—reflecting the Landau–Pomeranchuk–Migdal (LPM) effect—is small. This behavior contrasts with the $1/\theta$ scaling from the LT contribution, in agreement with the findings of \cite{Andres:2024xvk} based on the light-ray operator product expansion.

\subsection{Jet EECs in  p+Pb collisions}
\label{sec-pA}
In p+Pb collisions, besides CNM effects, possible nuclear effects can also be caused by comovers—secondary particles from p+A interactions unrelated to hard scattering. The comover region is confined to the transverse area of the p+Pb interaction. Its transport coefficient can be estimated from the charged multiplicity as a proxy for entropy density, $\hat{q}_{\text{co}}/\hat{q}_{\text{hot}}  = \frac{\rmd N_{ch}^{pPb}}{(R^{p}_{rms})^2\rmd\eta} / \frac{\rmd N_{ch}^{PbPb}}{(R^{Pb}_{rms})^2\rmd\eta}$, giving $\hat{q}_{\text{co}} \approx 0.6~\mathrm{GeV}^2/\mathrm{fm}$, where $R_{rms}$ are root-mean-square radii of proton or lead nucleus\cite{Fu:2024pic}. For a gluon jet with momentum $p_T$ and rapidity $y$, the energy-weighted differential cross section ${\rmd \Sigma_{\text{pPb}}^{\text{LT+NLT}}}/\rmd y \rmd p_T^2 \rmd\theta$ takes the same form as in the e+A case, with the kernel $\mathcal{K}_{\textrm{NLT}}(\theta,p_T)$ involving the gluon–gluon splitting function $P_{gg}(z)$. The path-length integral is then separated into comover and CNM regions, each characterized by $\hat{q}_{\text{co}}$ and $\hat{q}_{\text{CNM}}$.

\section{Results}
\label{sec:results}
\begin{figure}[h]
    \centering
\includegraphics[width=\linewidth]{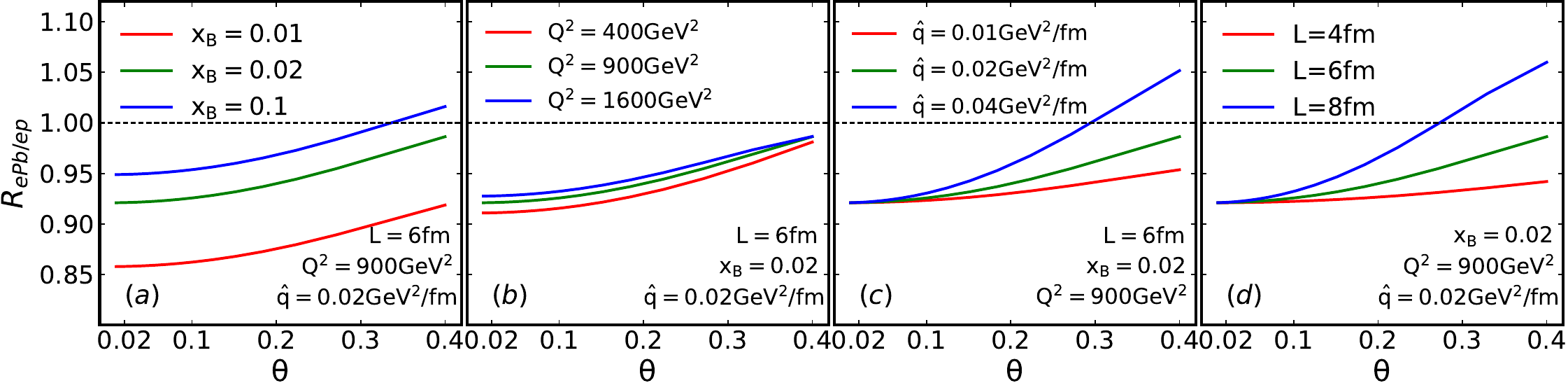}
    \caption{Ratio of the EEC in e+Pb to e+p as a function of the splitting angle $\theta$, obtained by varying each of the four parameters $x_B$, $Q^2$, $\hat{q}$, and $L$ while keeping the others fixed.}
    \label{fig::Ratio_eA}
\end{figure}

To study medium modifications of the jet EEC in DIS, we consider the ratio of the energy-weighted cross section in e+Pb to that in e+p collisions, 
$R_{\text{ePb}/\text{ep}}(\theta) = \frac{\rmd \Sigma^{\text{LT+NLT}}/\rmd x_B \rmd Q^2 \rmd\theta}{\rmd \Sigma^{\text{LT}}/\rmd x_B \rmd Q^2 \rmd\theta}$. 
Numerical results for different parameter choices are shown in Fig.~\ref{fig::Ratio_eA}. Initial-state effects suppress the EEC ratio at small angles, while final-state effects enhance it for $Q\theta > 2.4~\text{GeV}$. For $Q=30~\text{GeV}$, the region $\theta \lesssim 0.1$ is dominated by nonperturbative physics. Fig.~\ref{fig::Ratio_eA}(a) shows that varying $x_B$ only shifts the baseline without altering the $\theta$-dependence. In Fig.~\ref{fig::Ratio_eA}(b), we vary the values of $Q^2$ while keeping the remaining parameters fixed. We observe that the final-state effect on the EEC ratio is more pronounced at lower $Q^2$. In Figs.~\ref{fig::Ratio_eA}(c,d), we study the sensitivity of the final-state nuclear effect on $\hat{q}$ and $L$. As expected, increasing values of either $\hat{q}$ or $L$ strongly enhance the final-state effect at fixed $Q$ and $x_B$, resulting in an increasingly strong growth of the EEC ratio for $\theta > 0.1$.

\begin{figure}[H]
\centering
\includegraphics[width=6.45cm]{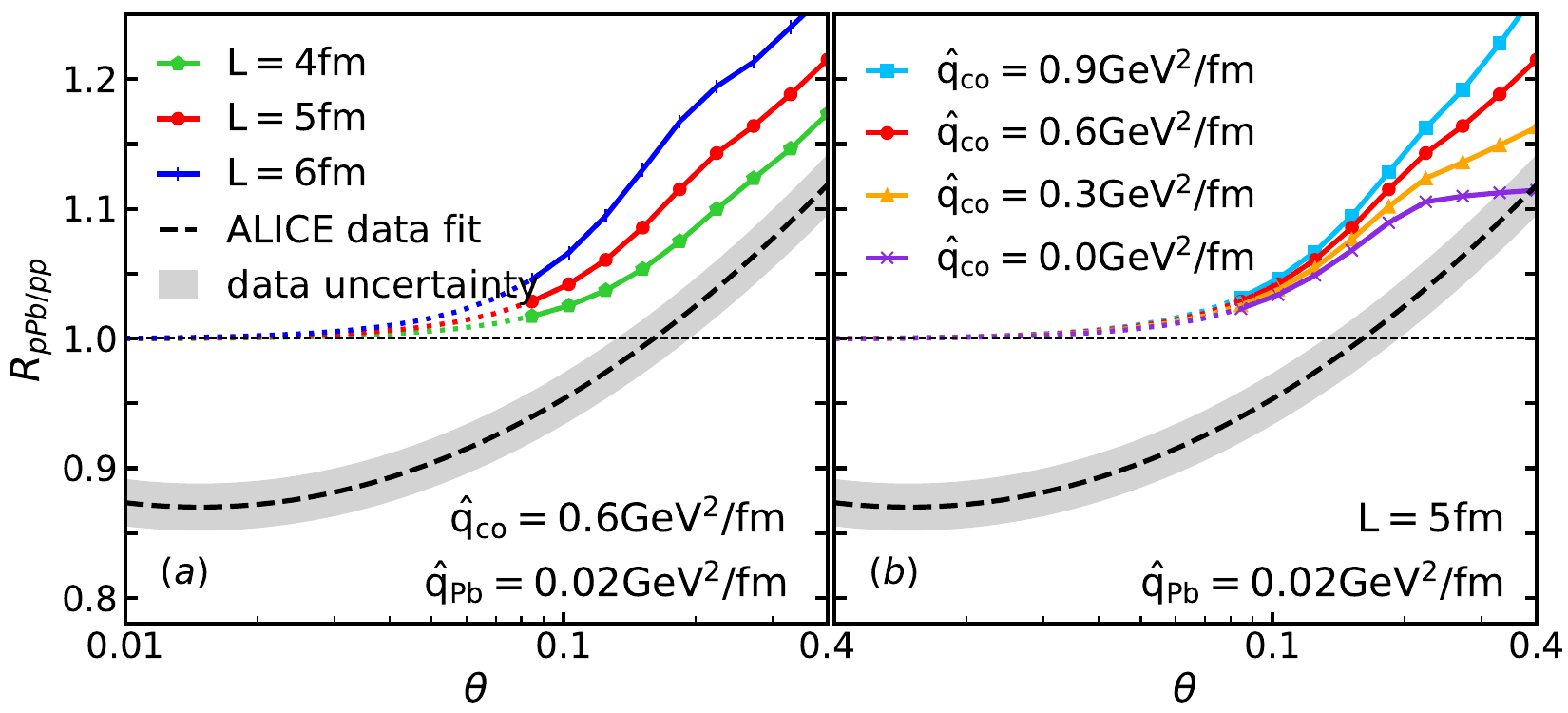}
\includegraphics[width=6.45cm]{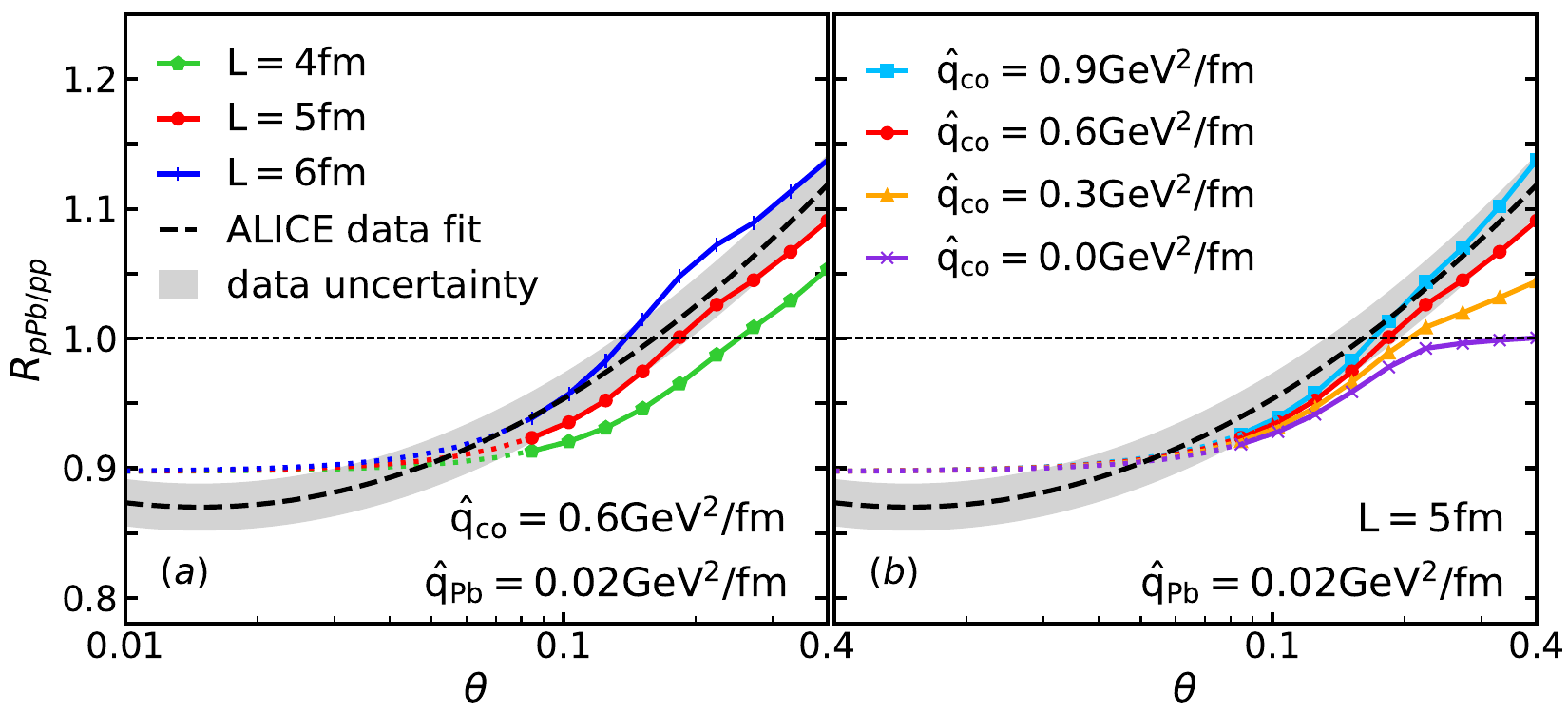}
\caption{Ratio of the EEC for a $30~\mathrm{GeV}$ jet at mid-rapidity in p+Pb to p+p as a function of $\theta$, compared with the ALICE data fit. In the panels (a,b), results are shown for variations of the path length $L$ (a) and $\hat{q}_{\mathrm{co}}$ (b). The panels (c,d) show the corresponding results with an overall multiplicative factor of $0.9$.}
\label{fig::Ratio_pA}     
\end{figure}

To estimate the nuclear modification of jet EECs in p+A collisions and compare with the ALICE measurement, we consider the ratio of normalized energy-weighted cross sections in p+A and p+p collisions, 
$R_{\text{pA}/\text{pp}}(\theta) = \frac{\sigma_{\text{pp}}}{\sigma_{\text{pA}}} \, \frac{\rmd \Sigma_{\text{pA}}^{\text{LT+NLT}} / \rmd y \rmd p_T^2 \rmd \theta}{\rmd \Sigma_{\text{pp}} / \rmd y \rmd p_T^2 \rmd \theta}$. 
For a gluon jet at midrapidity with $p_T=30$ GeV/c in p+p/Pb collisions at $\sqrt{s}=5$ TeV, this ratio is shown in the panels (a,b) of Fig.~\ref{fig::Ratio_pA} as a function of $\theta$, together with a fitted curve based on preliminary ALICE data\cite{Nambrath:2025qm} obtained by varying the path length $L$. The panel (a) illustrates the $L$-dependence for fixed $\hat{q}_{\text{co}}$ and $\hat{q}_{\text{Pb}}$, while the panel (b) shows the dependence on $\hat{q}_{\text{co}}$ for fixed $L$. Our perturbative calculation is applicable only for $\theta \gtrsim 0.08$, corresponding to the solid curves in Fig.~\ref{fig::Ratio_pA}.  The results indicate that the EEC ratio exceeds unity in this region: rescattering of intrajet partons with the medium enhances the splitting probability, leading to nuclear enhancement of the EEC. The effect grows with increasing path length $L$, and a larger comover density further strengthens the enhancement, yielding a nearly linear angular dependence due to the short comover path length. If one extrapolates the curves (without rigorous justification) into the nonperturbative region $\theta < 0.08$, shown as dotted lines, the ratio tends toward unity—consistent with the cancellation of nPDF effects in the definition of $R_{\text{pA}/\text{pp}}$.

By choosing appropriate values of the path length and the comover transport coefficient, our calculation of the EEC ratio reproduces the qualitative trend seen in the ALICE data. Quantitatively, however, the result overshoots the measured magnitude by about 10\%(A simple rescaling with a multiplicative factor of 0.9 yields excellent agreement with the data, as shown in the panels (c,d) of Fig.~\ref{fig::Ratio_pA}.). The origin of the overall discrepancy with the ALICE data remains unclear and warrants further study, including a careful examination of EEC normalization, potential jet selection biases, and the role of multiple scattering in the medium.

\section{Summary}
\label{sec:summary}
We present the complete calculation of nuclear modifications of the jet EEC in e+A collisions at NLT. Nuclear effects arise from both initial-state interactions, encoded in nuclear PDFs, and final-state interactions of the showering jet with the cold nuclear medium, described by in-jet transverse momentum broadening and the LPM effect through $\hat{q}L$. While nPDF induces a global modification across angles, rescattering in the medium enhances the EEC at $\theta \gtrsim 0.1$. We also apply the HT formalism to jets in p+A collisions and compare with preliminary ALICE data. Although the overall magnitude is larger than observed, the calculation reproduces the characteristic rise of the EEC ratio at large angles, driven by final-state rescattering. Our study highlights that the EEC is more sensitive to nuclear PDFs at small angles and to medium-induced broadening at large angles, making it a promising observable for extracting the nuclear modification factor for PDFs and $\hat{q}$ at the EIC, as well as for probing additional effects such as comovers in p+A collisions.
\\

\!\!\!\!\!\!\!\!\!\textbf{Acknowledge}: This work was supported by the grant DE-FG02-05ER41367 from the U.S. Department of Energy, Office of Science, Nuclear Physics. C.S. is supported in part by the National Science Foundation under grant numbers ACI-1550225 and OAC-2004571 (CSSI:X-SCAPE) within the framework of the JETSCAPE collaboration.

\end{document}